\documentstyle[11pt,psfig]{article} 
\newcommand{{\nc}}{\newcommand}
\nc\QED{\hfill \framebox[3mm]{}} 
 
\begin{document} 
 
\title       { Remarks on Bounds for Quantum codes} 
\author      {Alexei Ashikhmin\\ 
{\small Los Alamos National Laboratory }}
  \date{} 
\maketitle

\begin{abstract}
 We present some  results that show that bounds from classical coding theory still  work in many cases of quantum coding theory. 
 \end{abstract} 
  
\section{ Nondegenerate Quantum Codes} 

Let $Q$ be an $((n,K,d))$ nondegenerate quantum code with  projection operator $P$.
Let $A_i$ and $B_i$ be the  quantum enumerators of $Q$,  \cite{ref 1},\cite{ref 2}
$$
A_i=\sum_{wt(E)=i} \mbox{Tr}(EP)\mbox{Tr}(EP),
$$
$$
B_i=\sum_{wt(E)=i} \mbox{Tr}(EPEP),
$$
where $E$ is an error operator. (See \cite{ref 1},\cite{ref 2} for
definition of error operators and their weights.)

It is shown in \cite{ref 1},\cite{ref 2} that  
$$
A_t=\frac{1}{2^n} \sum_{i=0}^{n} B_i P_t (i,n),
$$
where $P_t(i,n)=\sum_{j=0}^{i} (-1)^j 3^{t-j} {i\choose j} {n-j \choose t-j}$ is the Krawtchouk polynomial. Replace $A_i$  by $A_i/K^2$ and $B_i$ by $B_i/K$. Then we get
\begin{equation}
A_t=\frac{1}{2^nK} \sum_{i=0}^{n} B_i P_t (i,n) \mbox{  and  }A_0=B_0=1.
\end{equation}
Now we can use well known techniques to get upper bounds for the quantity $2^nK$ (see for example 
\cite{ref 5}).We describe this technique here for completeness.

Let $f(x)=\sum_{i=0}^{n} a_i x^i$ be a polynomial . Consider its expansion by Krawtchouk polynomials
\begin{equation}
f(x)=\sum_{i=0}^{n} f_i P_i(x,n).
\end{equation}
Then from (1) and (2)  we have
$$
\sum_{i=0}^{n} B_i f(i)=\frac{1}{2^nK}\sum_{i=0}^{n} A_i f_i.
$$
Suppose that 
\begin{equation}
f_0>0, f_i \ge 0, \mbox{ if }  i=1,\ldots , n, \mbox{ and}
 f(0)>0, f(i)\le 0, \mbox{ if }
i=d, \ldots ,n.
\end{equation}  
Then taking into account that for nondegenerate codes $A_i=B_i=0$ for $ i=1, \ldots , d-1$,
we have 
\begin{equation}
2^nK=\frac{f(0)+\sum_{i=1}^{n} f(i) B_i}{f_0+\sum_{i=1}^{n} f_i A_i}\le\frac{f(0)}{f_0}.
\end{equation}
From classical coding theory we know many useful polynomials that satisfy conditions (3).
Here we list some of them.

Let $f(x)=4^{q-d+1}\prod_{j=d}^{n} (1-\frac{x}{j})$ \cite{ref 6}. Then
$$
f_i=\frac{{n-k \choose d-1}}{{n \choose d-1}}\ge 0.
$$
So we get a quantum Singleton bound \cite{ref 7}\cite{ref 8}: $K\le 2^{n-2d+2}$. Note that  $f_i=0$ for $i\ge n-d+2$ and hence
if a quantum code meets the Singleton bound it should be 
nondegenerate up to $n-d+2$ (a quantum code is called nondegenerate up to $t$ if $A_i=0$ for $i=1, \ldots, t-1$
\cite{ref 8}.)(In \cite{ref 6} this proposition is proven for all quantum codes.)
\par
Let $d=2e+1$ and 
$$
f_i=\left ( \frac{P_e(i-1,n-1)}{\sum_{i=0}^{e} 3^i {n \choose i}}\right )^2.
$$
Then \cite{ref 6} $f(i)=0$ for $i=d,\ldots,n$ and $f(0)=4^n/\sum_{i=0}^{e} 3^i {n \choose i}$, and 
from (4) we  get the quantum analog of the Hamming bound: $K\le 2^n \sum_{i=0}^{e} 3^i {n \choose i}$.

Analogously, from \cite{ref 5} it is straightforwardly follows that
$$
K\le L^{n}(d)/2^n,
$$
where
$$
L^{n}(x)=\left \{
\begin{array}{lcc}
L_k^n(x), & \mbox{ if } & d_k(n-1)+1<x<d_{k-1}(n-2)+1\\
4L_k^{n-1}(x), & \mbox{ if } & d_k(n-2)+1<x<d_{k}(n-1)+1.\\  
\end{array} 
\right.
$$

$$
L_k^n(x)=\sum_{i=1}^{k-1}{n \choose i}3^i-{n \choose k}3^k\frac{P_{k-1}(x-1,n-1)}{P_k(x,n)},
$$
and $d_k(n)$ is the smallest root of $P_k(x,n)$.
 
Using the asymptotic version of this bound \cite{ref 5},  as $n$ tends to infinity
\begin{equation}
\label{lev}
\log_2 K \ge 2 H_4(x)-n/2 \mbox{ and } d < n \gamma_4(x),
\end{equation}
where 
$$
\gamma_q(x)=1/q((q-1-(q-2)x-2\sqrt{(q-1)x(1-x)}), \mbox{ and } 
$$
$$
H_q(x)=-x\log_qx-(1-x)\log_q(1-x)+x\log_q(q-1).
$$
(See figure 1,B.) Note that as $\frac{log_2K}{n}$  tends to zero $\frac{d}{n}\approx 0.316$ which is better then the value $n/3$
from \cite{ref 12}. However the  latter bound is valid for all quantum  codes. 

\section{Stabilizer Codes}
\subsection{Nondegenerate Stabilizer Codes}

In the case of nondegenerate stabilizer codes just all bounds for classical codes
are valid. For example we can use the best known asymptotic bound for nonbinary codes from \cite{ref 14}. (See figure 1, A). Note that as $\frac{log_2K}{n}$  tends to zero $\frac{d}{n}\approx 0.308$

\subsection{General Case}

Let $Q$ be a  quantum stabilizer $[[n, k]]$ code (degenerate or nondegenerate).  In \cite{ref 10} it was shown that if the minimum distance of $Q$ is $d$ then there exist a
classical binary code $[n-1,k,d]$. Using this observation and  a classical linear programming bound, an upper bound for quantum stabilizer codes was obtained \cite{ref 10} (see figure 1,C). Arguments used in \cite{ref 10} seem to be complicated. We show that a simpler approach yield to somewhat stronger bound. 
\par
Let $C$ and $C^{\perp }, C\subseteq C^{\perp }$, be additive $[n,n-k]$ and $[n,n+k]$ codes over $\mbox{GF}(4)$ associated with  $Q$. If an additive code has cardinality 
$2^t$ we will say that $t$ is its dimension. So the dimensions of $C$ and $C^{\perp }$ are
$n-k$ and $n+k$ respectively. 
It is known that $Q$ has minimum distance $d$ iff all code words from the set $C^{\perp }\setminus C$
have weights greater than or equal to $d$ 
(see \cite{ref 13} for details). We can consider codes $C$ and $C^{\perp }$ as linear binary 
codes $\cal{C}$ and $\cal{C}^{\perp }$ of length $2n$. $\cal{C}^{\perp }$ is dual to $\cal{C}$ with respect to the symplectic inner product, i.e. if $(\bf{a},{\bf b})\in \cal{C}$ and $(\bf{a}',{\bf b}')\in\cal{C}^{\perp }$ 
then $({\bf a},{\bf b})\bullet({\bf a}',{\bf b}')=<{\bf a},{\bf b}'>\oplus <{\bf a}',{\bf b}>=0$.
 In \cite{ref 10} it was shown that a generator matrix $G$
of $\cal{C}$ can be written in the form
$$
G=\left [   
\begin{array}{cccccc} 
I_s& A_1 & A_2 & B_1 & B_2 & 0  \\
0&0&0& D_1&D_2&I_r  \\
 \end{array} 
\right ], 
$$
 where $I_s$ and $B_1$ are $s \times s$ matrices, $A_1$ and $B_1$ are $s \times k$ matrices,
$A_2$ and $B_2$ are $s \times r$ matrices, $D_1$ is $r \times s$ matrix, and $D_2$ is $r \times k$
matrix. It is not difficult to check that all rows of the matrix $G^{\perp }$
$$
G^{\perp }=\left [ 
\begin{array}{cccccc} 
I_s& A_1 & A_2 & B_1 & B_2 & 0  \\
0&0&0& D_1&D_2&I_r\\
0&0&0&A_1^{\top} & I_k& 0 \\
0&I_k& D_2^{\top} &B_2^{\top} & o& 0   \\
\end{array} 
\right ],
$$
are linearly independent and orthogonal (with respect to the symplectic inner product) to
rows of the matrix $G$. So the matrix $G^{\perp }$ is a generator matrix for the code $\cal{C}^{\perp }$.

It is obvious that if the minimum distance of codewords from $C^{\perp }\setminus C$ is $d$ then
the minimum distance of the binary code, say $S$,  with generator matrix 
\begin{equation}
\left [ 
\begin{array}{cccccc} 
 0&0&0&A_1^{\top} & I_k& 0\\
0&I_k& D_2^{\top} &B_2^{\top} & o& 0\\
\end{array} 
\right ]
\end{equation}
 is not less than $d$. The code $S$ is an $[n+k,2k]$ binary code. Applying now any classical  bound 
for this code we get a bound for code $Q$.
For example we can use the second McEliece-Rodemich-Rumsey-Welch bound \cite{ref 11}. The  resulting quantum bound is
better than the bound from \cite{ref 10} (see figure 1, D). In fact in \cite{ref 10} only a code with the generator 
matrix 
$$
\left [ 
\begin{array}{cccccc} 
 0&0&0&A_1^{\top} & I_k& 0\\
 \end{array} 
\right ]
$$
was taken into account.  

We can get a stronger bound if we consider codes $C$ and $C^{\perp}$ over $GF(4)$.
A generator matrix of $C$ can be written in the form \cite{ref 13}
\begin{equation}
G=\left[
\begin{array}{ccc}
I_{k_0} & \omega A_1 & B_1 \\
\omega I_{k_0} & \omega A_2 & B_2 \\
        & I_{k_1} & A_3 \\
\end{array}
\right ],
\end{equation}
where $B_j$ is a binary  matrix, $A_j$ is an arbitrary matrix, and $\omega $
is an element of $\mbox{GF}(4)$ of order 3.  We will say that $G$ defines  a code of type $4^{k_0}2^{k_1}$. If in some $l$ coordinates an  additive code,
say $S$, of length $n$ contains only $0$-s and $\alpha$-s,
$\alpha \in \mbox{GF}(4),\alpha\not =0$, then we will say that $S$ is a mixed code of lengths $l$ and $n-l$. Note that if $C$ is a code of type $4^{k_0}2^{k_1}$ of length $n$ then corresponding quantum code $Q$ is 
an $[[n,k]]=[[n,2n-2k_0-k_1]]$ code.
  
\newtheorem{proposition}{Proposition}
\begin{proposition}
Let $d$ be a minimum distance of $Q$. Then $d$ is not grater than the minimum distance 
of the best additive mixed code of lengths $k_1$ and  $n-k_0-k_1$  and 
dimension $2n-4k_0-2k_1$. In particular 

i) if $k_1=0$ then $d$ is not greater
than the minimum distance of the best $[n-k_0,2n-4k_0]=[(n+k)/2, 2k]$ additive code. 

ii) if $k_1< \frac{2n-4k_0}{3}=2k$ 
then $d$ is not greater than the  minimum distance
of the best $[n-k_0-k_1,2n-4k_0-3k_1]=[\frac{n+k-k1}{2},2k-k_1]$ additive code.


\end{proposition}

{\bf Proof}

 Let $G'$ be a generator matrix of a complementary code of $C$ to $C^{\perp}$,
i.e. $\left[ {G \atop G'}\right]$ is a generator matrix of the code $C^{\perp}$.
Due to the structure (7) of matrix $G$ we can make elements of $G'$ on the first $k_0$ positions  equal $0$ and elements on the next $k_1$ positions equal to $0$ or $\omega$.  So $G'$ will have the form$[0~D_1~D_2]$ where   $D_1$ is an $2n-4k_0-2k_1\times k_1$
matrix consisting from $0$-s and $\omega$-s and $D_2$ is is an arbitrary $2n-4k_0-2k_1\times n-k_0-k_1$ matrix. Since the minimum distance of a complementary code
has to be not less than minimum distance of the quantum code $Q$ we get 
the assertion. 

i) follows from the previous.

ii) In the case $k_1< \frac{2n-4k_0}{3}$ the matrix $[0~D_1~D_2]$ can be transformed  to the form  $$
\left[
\begin{array}{ccc}
0 & A_1 & B_1 \\
0 &  0  & B_2
\end{array}
\right],
$$
where $A_1$ is an $k_1\times k_1$ matrix consisting form $0$-s and $\omega$-s,
$B_1$ and $B_2$ are arbitrary  $k_1\times n-k_0-k_1$ and $2n-4k_0-3k_1\times n-k_0-k_1$ matrices.
Since the subcode with the generator matrix $[0~0~B_2]$ has length $n-k_0-k_1$, 
dimension $2n-4k_0-3k_1$ and its minimum distance has to be not less than minimum distance 
of $Q$, the assertion follows.  \QED


Let $S$ be a mixed code  of lengths $l$ and $n-l$ 
and dimension $k$ . 



Using standard arguments we can get the Hamming bound for the code $S$.

\newtheorem{lemma}{Lemma}
\begin{lemma}
$$ 
\sum_{i=0}^{e} \sum_{j=0}^{i}{l \choose j}3^{j-i}{n-l \choose i-j} \le 2^{2n-l-k},
$$
were $e=\left\lceil \frac{d-1}{2} \right\rceil$.
\end{lemma}


Combining Lemma 1 and Proposition 1 we get a bound of Hamming type for degenerate codes.

\begin{proposition}

$$ 
\sum_{i=0}^{e} \sum_{j=0}^{i}{k_1 \choose j}3^{j-i}{n-k_0-k_1 \choose i-j} \le 2^{2k_0+3k_1}=
4^{\frac{n-k}{2}+k_1},
$$
were $e=\left\lceil \frac{d-1}{2} \right\rceil$.
\end{proposition}

Consider the case $k_1=0$. Then we have 
$$ 
\sum_{i=0}^{e} 3^{i}{n-k_0\choose i} \le 
4^{\frac{n-k}{2}}.
$$
Asymptotically  this becomes 
$$
4^{\frac{n+k}{2}}H_4(\frac{d}{n+k})\le 4^{\frac{n-k}{2}}
$$
$$
\lambda=\frac{k}{n}\le \frac{1-H_4(\mu )}{1+H_4(\mu ) };~~ \delta=\frac{d}{n}\ge \left(1+\frac{1-H_4(t)}{1+H_4(t)}\right)t.
$$

In the case $k_1=0$ we can also apply asymptotic bounds for complementary codes $[(n+k)/2,2k]$ from Proposition 1.
For example we can use an asymptotic bound from \cite{ref 14} (see fig. 1,E). Note that as $\frac{log_2K}{n}$  tends to zero $\frac{d}{n}= 0.375$.

Let $k_1>0$. Then according to Proposition 1 we can estimate a minimum distance of $Q$ for $n-k_1 < k < k_1/2$ as minimum distance of an additive $[\frac{n+k-k_1}{2},2k-k_1]$ code. For example using 
the asymptotic bound from  \cite{ref 14} and fixing $k_1=0, k_1=0.2, k_1=0.5$ we get bounds A,B,C, figure 2. 

If $k>k_1/2$ we have to apply bounds for a mixed code. Generalizations of asymptotic 
bounds for mixed codes are not known to the author.     
Note that we  can still  use asymptotic bounds  based on representation of complimentary code in the binary form (6) or the bound from [7] 
(bounds C and D on fig.1)
for an arbitrary quantum stabilizer     
code.   
 

\begin{figure}
\centerline{\psfig{file=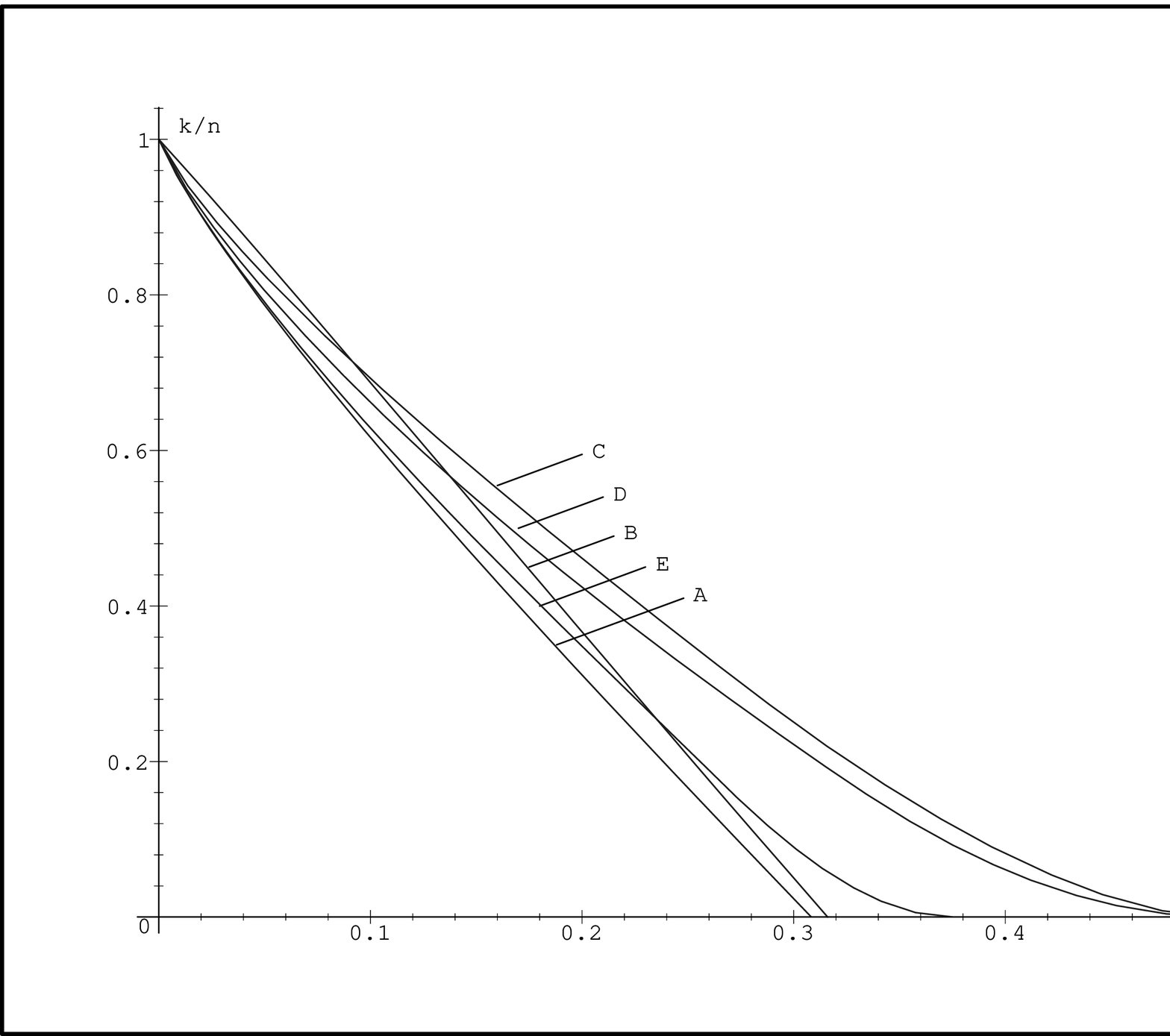,height=18cm}}
\caption{\em  A is a bound for nondegenerate stabilizer codes; B is a bound for degenerate codes;
C is a bound for degenerate stabilizer codes from  [7]; D is bound based on a complementary 
code written in the binary form (6); E is a bound for degenerate stabilizer codes based on the classical bound from 
[11]. } 
\label{ref.fig1} 
\end{figure}

\begin{figure}
\centerline{\psfig{file=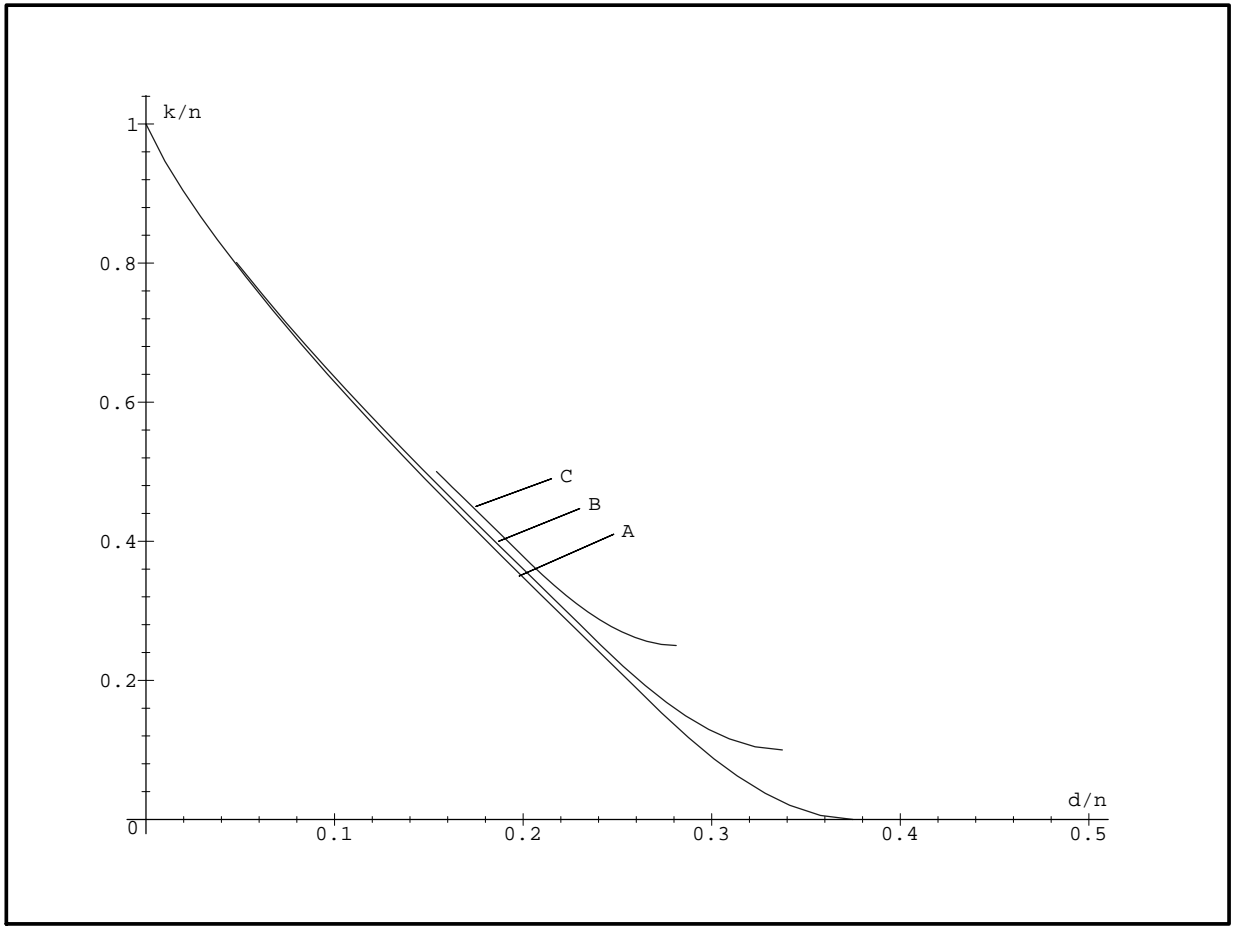,height=18cm}}
\caption{\em  A is a bound for $k_1=0$; B is a bound for  $k_1=0.2$;
C is a bound for  $k_1=0.5$.  } 
\label{ref.fig2} 
\end{figure}

\end{document}